\documentclass   [a4paper,12pt]{article}
\usepackage{graphicx}
\begin{document}
\begin{center}
{\bf\large The velocity increase of mass and the classical   
               physics} \\
    Milo\v{s} V. Lokaj\'{\i}\v{c}ek, Institute of Physics, v.v.i. \\
     Academy of Sciences of the Czech Republic, 18221 Prague \\
                     e-mail: lokaj@fzu.cz
\end{center}

Abstract

In the past century it was believed that both the main theories (quantum mechanics and special relativity) predicted the existence of physical processes that could not be explained in the framework of classical physics. However, it has been shown recently that the solutions of Schr\"{o}dinger equation have described the physical situation practically in full agreement with classical equations. The given equation represents the combination of classical equations with the statistical distribution of corresponding parameters and the properties of microscopic objects may be interpreted on the ontological basis as it corresponds to our sensual knowledge.
 
 It will be shown now that also the main experimentally relevant relativistic phenomenon (i.e., the mass increase with velocity) may be interpreted in the framework of classical physics. A different prediction for this increase will be then derived, which gives the possibility to decide on experimental basis which alternative is more preferable (relativistic or classical). \\[1cm]

The physics of the twentieth century tried to convince the human community that the laws ruling in microscopic nature differed from the classical ones in a decisive way. However, it has been shown recently (see \cite{conc1,adv} and papers quoted there) that the Copenhagen quantum mechanics has been based on some unphysical assumptions and the statistical (or ensemble) alternative (see, e.g., \cite{home}) being described practically by the mere Schr\"{o}dinger equation has led to the results fully equivalent to equations of classical physics, to which the statistical distribution of some parameters has been added. It is in full agreement with the recent results of U. Hoyer \cite{hoyer}. There is not any contradiction to the ontological sensual knowledge, either.

And it is quite natural to ask how it is with the "non-classical" phenomena predicted by special theory of relativity. It is practically evident that the most of relativistic phenomena may be hardly tested experimentally in a direct way. There is in principle the only prediction, i.e., the mass increase with rising velocity that is fully experimentally relevant. It means that the original classical relation between the force $F$ and the acceleration $a$ 
\begin{equation}
                     F = m_0\, a      \label{clas}
\end{equation}
must be modified. It must be substituted by more general relation
\begin{equation}
                     F = M_v a_v               \label{gen}
\end{equation}
where $M_v$ rises with rising $v$; and $a_v$ diminishes correspondingly. This fact has been confirmed with the help of different particle accelerators in a qualitative way. However, it has been never shown explicitly that this increase follows the corresponding relativistic formula.

We will show now that the given mass increase is not a specifically relativistic property, but that it may be brought to agreement with classical ontological picture. However, at the same time it will be shown that the velocity dependencies will be different in relativistic and classical alternatives, which gives also the possibility of deciding the question on experimental basis. 

Let us introduce now two assumptions that might represent the bases in both the alternatives (classical and relativistic): 

 - energy $E_v$ and mass $m_v$ that fulfill the condition $E_v=m_vc^2$ ($c\;$ being a constant parameter having the dimension of velocity) may be attributed to any object moving with velocity $v$;  
 
 - the mass of an object rises as $m_v=m_0f(v/c)$ where $f(\beta)$ is a monotony rising function (to be determined); it holds $f(0)=1$. 
 
Let us assume further that the ratio between quantities $F$ and $a_v$ is given by Eq. (\ref{gen}). The force $F$ acting in the direction of movement gives the energy increase $dE=Fds$ where $ds$ is the corresponding track element. And as it holds $vdv=a_vds$ for accelerated motion it is possible to write 
\begin{equation}
         M_v=\frac{1}{v}\frac{dE_v}{dv} = 
 \frac{m_0}{\beta}\frac{df(\beta)}{d\beta}, \;\;\;\beta=\frac{v}{c}.
\end{equation}

Parameter $M_v$ characterizes the resistance of the moving object against the velocity change and its value depends on the velocity $v$ of the given object. And it is natural to try first to extend the classical relation ({\ref{clas}) and to assume that the resistance against motion is given by corresponding value of mass, i.e., 
\begin{equation}
                     M_v = m_v  \;.      \label{clas2}
\end{equation}
One obtains thus the condition for determining the function $f(v/c)$; it holds then
\begin{equation}
             f(\beta) = e^{\frac{1}{2}\beta^2}\; ,   \label{fb1}
\end{equation}
which is, of course, quite different from the expression derived in special relativity theory.
  
In the relativity theory the different dependence is being made use of
\begin{equation}
                   f(\beta)= \frac{1}{\sqrt{1-\beta^2}}   \label{fb2}
\end{equation}
where $c=v/\beta$ is equal to light velocity $c_l$. In contradistinction to condition (\ref{clas2}) one obtains then
\begin{equation}
                  M_v = \frac{m_0}{(1-(\frac{v}{c})^2)^{3/2}} \; .          
\end{equation}
Introducing
\begin{equation}
                   m^{(r)}_v = \frac{m_0}{\sqrt{1-\beta^2}}
\end{equation}
it is possible to write
\begin{equation}
          M_v=\frac{d}{dv}( m^{(r)}_v v) \;. 
\end{equation}
And if holds for the effect of the force
\begin{equation}
         F=\frac{d}{dt}(m^{(r)}_v v) \;;
\end{equation}
or: the time change of momentum is equal to the corresponding force value.

We have, therefore, two different (classical and relativistic) formulas (\ref{fb1}) and (\ref{fb2}) characterizing the effect of the force in the dependence on the velocity of a moving object. The increases of the mass with velocity in individual cases are significantly different. For the dependence between the velocity and energy one can write in the first case
\begin{equation}
                 v = c \sqrt{2 lg(\frac{E_v}{E_0})}
\end{equation}
and in the other case                   
\begin{equation}
                v = c \sqrt{1-(\frac{E_0}{E_v})^2} \;.
\end{equation}
It means that in the classical case the velocity of a matter object rises permanently, but slowly (logarithmically) with the rising energy, while in the relativistic case it must come quickly near to the limit velocity $c$ that is represented now by light velocity $c_l$. The corresponding dependencies are represented in Fig. 1; the object energy in GeV is shown on the horizontal axis and the ratio $\beta = v/c$ is given on the vertical axis. The full line represents the relativistic case; it holds $c=c_l$. The other two dependencies correspond then to the classical alternative: it has been chosen $c=c_l$ for the upper case, and $c = c_l/3$ for the lower one. The two alternatives (relativistic and classical) exhibit, therefore, different characteristics. And it may be easily decided between them on the basis of corresponding experimental data; and eventually, the numerical value of parameter $c$ may be established in the classical case.  
\\[5mm]

\begin{figure}[htb]
\begin{center}
\includegraphics*[scale=.45, angle=-90]{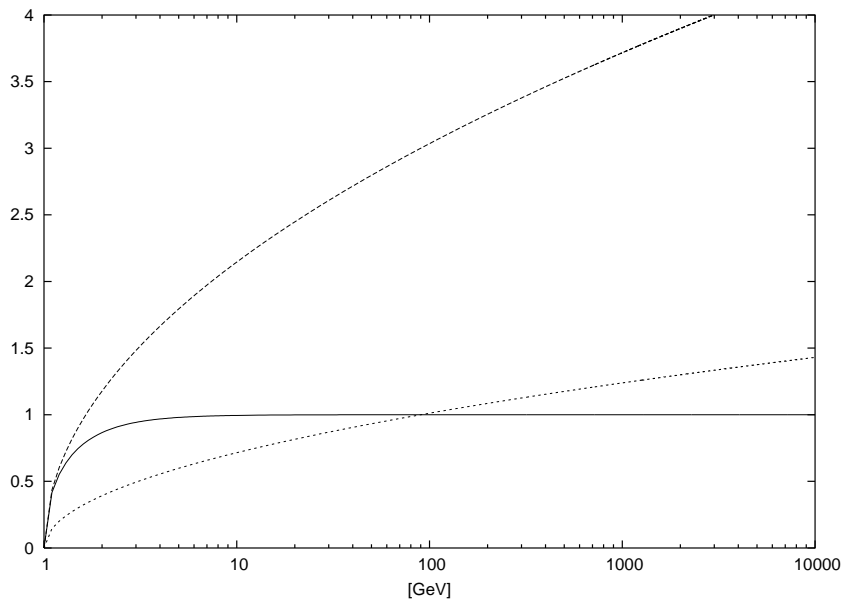}
  \caption { \it { The velocity increase in the dependence on the energy of moving object with rest mass 1 GeV; the energy in GeV is shown on the horizontal axis, ratio $\beta=v/c$ on vertical axis. Full line - behavior according to relativity theory, dashed lines - two different possibilities from continuous classical set (see text).    } }
 \end{center}
 \end{figure}

Even if the decision between the two possibilities must be given on the experimental basis the increasing resistance against motion change may be well understood in the framework of classical ontological interpretation. However, some new questions would be opened: Is the change of resistance against motion accompanied by a change of internal structure of a moving object or not? And what is really represented by the quantity $m_v$? And further: How is the force being transmitted to an object in "physical vacuum"? However, first the decision between different alternatives on experimental grounds should be done.

{\footnotesize
 
\vspace{4mm}

{\bf\small Addendum}\\
\footnotesize The preceding manuscript was submitted to Physical Review Letters on Nov. 2, 2007. However, its publication has been immediately refused; see the following e-mail obtained on Nov. 5, 2007: \\ 

From prl@aps.org Wed Nov  7 08:56:24 2007 \\
Date: Mon, 5 Nov 2007 08:11:00 -0500 (EST) \\
From: prl@aps.org  \\
To: lokaj@fzu.cz  \\
Subject: Your manuscript LL10880 Lokajicek  \\

Re: LL10880 \\
    Velocity increase of mass and classical physics  \\
    by Milo\v{s} V. Lokaj\'{\i}\v{c}ek \\

Dr. Milos V. Lokajicek  \\
Institute of Physics \\
Academy of Sciences of CR  \\
Na Slovance 2  \\
18221  Prague  \\
CZECH REPUBLIC  \\

Dear Dr. Lokajicek, \\
Your manuscript has been considered.  We regret to inform you that we \\
have concluded that it is not suitable for publication in Physical  \\
Review Letters.  \\

Yours sincerely, \\
Jerome Malenfant  \\
Senior Assistant Editor  \\
Physical Review Letters   \\
Email: prl@ridge.aps.org  \\
Fax: 631-591-4141  \\
http://prl.aps.org/  \\

\end{document}